\documentclass[fleqn,usenatbib]{mnras}
\usepackage{newtxtext,newtxmath}
\usepackage[T1]{fontenc}
\DeclareRobustCommand{\VAN}[3]{#2}
\let\VANthebibliography\thebibliography
\def\thebibliography{\DeclareRobustCommand{\VAN}[3]{##3}\VANthebibliography}
\usepackage{graphicx}	
\usepackage{amsmath}	
\title[Calibration and applications of the all-sky cameras]{Calibration and Applications of the All-Sky Camera at the Ali Observatory in Tibet}
\author[J. Yin et al.]{
Jia Yin,$^{1}$\thanks{E-mail: jyin@bao.ac.cn (JY); liuly@nao.cas.cn (LYL)}
Yongqiang Yao,$^{1}$
Xuan Qian,$^{1}$
Liyong Liu,$^{1}$\footnotemark[1]
Xu Chen,$^{1}$
and Liuming Zhai$^{1}$
\\
$^{1}$National Astronomical Observatories, Chinese Academy of Sciences, 20A Datun Road, Chaoyang District, Beijing, 100101, China
}

\date{Accepted XXX. Received YYY; in original form ZZZ}
\pubyear{\the\year{}}

\begin{document}
\label{firstpage}
\pagerange{\pageref{firstpage}--\pageref{lastpage}}
\maketitle

\begin{abstract}
A high-precision calibration method for all-sky cameras has been realized using images from the Ali observatory in Tibet, providing application results for atmospheric extinction, night sky brightness, and known variable stars. This method achieves high-precision calibration for individual all-sky images, with the calibration process introducing deviations of less than 0.5\,pixels. Within a 70-degree zenith angle, the calibration deviation of the images is less than 0.25\,pixels. Beyond this angle, the calibration deviation increases significantly due to the sparser distribution of stars. Increasing the number of stars with zenith angles greater than 70\,degrees used for calibration can improve the calibration accuracy for areas beyond the 70-degree zenith angle, reducing the calibration deviation at an 85-degree zenith angle to 0.2\,pixels. Analysis of the all-sky images indicates that the atmospheric extinction coefficient at the Ali Observatory is approximately 0.20, and the night sky background brightness is about 21 magnitudes per square arcsecond, suggesting the presence of urban light pollution.
\end{abstract}


\section{INTRODUCTION}

All-sky images are captured by all-sky cameras equipped with fisheye lenses. The field of view (FOV) of these images can exceed 180\,degrees, readily recording atmospheric changes and astronomical events across the entire local sky. This provides significant advantages in both the temporal and spatial dimensions, making them widely used in astronomical observatories for assessing cloud cover \citep{2008SPIE.7012E..24S, 2014arXiv1402.4762M, 2020RAA....20..149X, 2024MNRAS.529.1195Q} and night-sky brightness \citep{2016JQSRT.181...33D, 2017AJ....154....6Y, 2018JQSRT.205..278H}, monitoring fireballs \citep{2008EM&P..102..241W, 2016pimo.conf...76G, 2019MNRAS.483.5166D, 2021MNRAS.506.5046V}, bright transients \citep{2005AN....326..428S, 2005PASA...22..111S} and bright variable stars \citep{2017AJ....154....6Y}, and researching airglow emissions \citep{1995GeoRL..22.2833T, 2018JGRA..123.9619Y}. However, due to the extremely large FOV of the all-sky images, distortion is quite pronounced, making the relationship between image coordinates and astronomical coordinates challenging to resolve. The relationship between image coordinates and astronomical coordinates is crucial for the studies mentioned above. It enables precise description of cloud coverage \citep{2005PASP..117..972S}, calculation of fireball trajectory and orbit parameters from observations at two or more sites \citep{2009MNRAS.392..367T, 2021MNRAS.508..326M}, and determination of the mesospheric wave structures through airglow emissions \citep{2013JASTP..93...21T}.

Based on the most fundamental zenith (or azimuth) projection, various models and methods have been developed for calibrating the distortion in all-sky images. \citet{2005PASA...22..111S} presents a fuzzy logic-based algorithm to estimate the required coordinate transformations. The algorithm can be considered as a complex interpolation method, using a set of reference stars to build a fuzzy logic model. Increasing the number of stars can enhance the accuracy of interpolation, but it also makes the fuzzy logic model more complex.

A widely used lens distortion model is provided by \citet{Brown1971} to calibrate the radial and decentering distortion of close-range cameras, and the intrinsic and extrinsic parameters of the camera model can be estimated from the planar object \citep{888718}. \citet{1642666} employs a more flexible approach to represent the complex distortion patterns of fisheye lenses, and provides a generic camera model and calibration method for conventional wide-angle and fisheye lenses. The calibration of a fisheye lens can be performed using an image of the planar object.  

For astronomical images, stars are more convenient reference points for model calibration. All-sky images, as a special class of astronomical images, can also be described according to the conventions of the Astronomical Image Processing System (AIPS). \citet{2002A&A...395.1061G} and \citet{2002A&A...395.1077C} describe a generalized method for assigning physical coordinates in World Coordinate System (WCS) to image pixels for all spherical map projections in astronomy. Meanwhile, the Simple Imaging Polynomial (SIP) convention is used for representing non-linear geometric distortion \citep{2004ASPC..314..551C, 2005ASPC..347..491S}. 

An astrometric model has been proposed to address the optical distortions of an all-sky camera, implemented through a parametric description that treats the alt-azimuth coordinates as a function of the image coordinates \citep{1987BAICz..38..222C, 1995A&AS..112..173B}. The astrometric parameter models have a high degree of complexity, and when the initial estimates are not accurate enough, parameter solving may encounter problems with local minima and crosstalk. \citet{2019A&A...626A.105B} adopts a new parametrisation to address these issues. Additionally, \citet{2022PASP..134c5002T} used the parameter model to obtain a star dataset, which was used for machine learning to achieve the transformation between image coordinates and celestial coordinates.   

To analyze the all-sky images, we have developed an image segmentation method that facilitates the automatic identification of stars and the calibration of the images. Section \ref{sec:monitor} describes the cloud monitor we use to obtain all-sky images. Section \ref{sec:method} introduces the all-sky camera models, provides a detailed description of the method for automatically calibrating the distortion in all-sky images, and discusses the calibration deviations. Section \ref{sec:application} provides some application results of the all-sky camera at the Ali Observatory. Finally, we provide the conclusion in Section \ref{sec:conclusions}.

\section{THE CLOUD MONITOR} \label{sec:monitor}
\begin{figure}
\centering
\includegraphics[width=.46\textwidth]{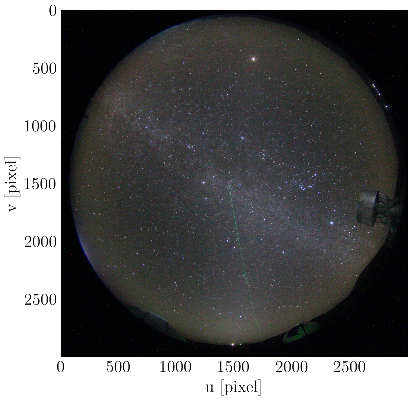}
\caption{A clear all-sky image is cutted based on the field of view.}
\label{fig:cut_img}
\end{figure}

The cloud monitor was designed to obtain cloud images of the whole sky and was employed at Ali Observatory \citep{2020RAA....20...84L} at an elevation of 5050\,m in Tibet, China. It contains a commercial Canon EOS~600D camera with a Sigma fisheye lens of F/2.8, and a shield for protection. The camera contains a CMOS chip with three channels (RGB), and the range of spectral sensitivity is from 400 to 900\,nm. The FOV of the camera and the image sizes are listed in Table~\ref{tab:para}. The effective coverage of the all-sky image was about 2800\,pixels in diameter, yielding an average plate scale of $0\fdg064$ per pixel, as shown in Fig.~\ref{fig:cut_img}. 

\begin{table}
\caption{The cloud monitor.}
\label{tab:para}
\begin{tabular}{l|c}
\hline
Parameter & Value\\
\hline
Camera                      &  Canon ESO 600D \\
Lens                        &  Fisheye lens of Sigma \\
Focus length (mm)           &  4.5 \\
CMOS                        &  RGB \\
Image size (pixel)          &  5184$\times$3456 \\
Field of view ($^{\circ}$)  &  180 \\
\hline
\end{tabular}
\end{table}

\subsection{Camera control}
The images captured by the Canon camera record exposure parameters (ISO, aperture, and exposure time) as well as evaluative metering values (the "MeasuredEV" and "LightValue" in the meta information). The relationship between them is as follows \citep{BILISSI2011227}: 
\begin{equation}
{\rm EV}_{\rm ISO} = \log_2\left(\frac{N^2}{t}\cdot \frac{100}{S}\right)
\end{equation}
where, $S$ stands for ISO, $N$ is the aperture value (f-number), $t$ is the shutter speed (or exposure time), and ${\rm EV}_{\rm ISO}$ is the evaluative metering value under ISO conditions. The evaluative metering values will vary with changes in ambient light, such as noticeable differences under various weather conditions during the day, at dawn and dusk, on moonlit nights, and on moonless nights. The "LightValue" is more responsive to changes in ambient brightness than "MeasuredEV", with its range varying from -9 to 16 from night to day. 

We have developed a camera control strategy, utilizing the digital camera application software gPhoto2\footnote{\url{http://gphoto.org/}} within the computer to control the camera for exposures taken at 10-minute intervals throughout the day and night. Each shooting session consists of two exposures: the first exposure is a test exposure, with default settings of $S=100, N=5.6, t=1/4000$, to prevent over-saturation of images caused by sunlight, during which the "LightValue" of the image is retrieved using the application ExifTool\footnote{\url{https://exiftool.org/index.html}}. The second exposure then adjusts the exposure parameters based on the "LightValue". The "LightValue" is divided into three segments at the boundaries of five and negative five. For each segment, fixed ISO and aperture settings are assigned, which are $S=100, N=5.6$, $S=1600, N=4.5$, and $S=6400, N=2.8$, respectively. The exposure time is adjusted to achieve coverage of the "LightValue", but the maximum exposure time is limited to 30 seconds. This enables the adjustment of exposure parameters in response to changes in ambient light levels, facilitating continuous, automatic cloud cover monitoring throughout the entire period. The camera's long exposure noise reduction feature has been enabled, allowing for automatic dark frame subtraction. As a result, the shooting time for exposures longer than one second will be doubled. The original images are saved in both CR2 and JPG formats.

\subsection{Artificial star experiments}
\begin{figure}
\centering
\includegraphics[width=.46\textwidth]{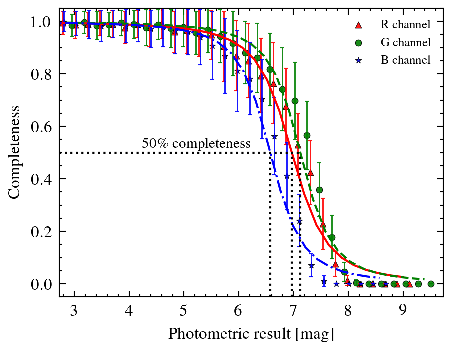}
\caption{The figure illustrates the results of the completeness test. The red triangles, green circles, and blue pentagons represent the average recovery probabilities of artificial stars for each magnitude bin in the R, G, and B channels, respectively. The corresponding error bars depict the standard deviations of these probabilities. The fitting relationships for the three channels, as described by \citet{1995AJ....109.1044F}, are shown by red solid lines, green dashed lines, and blue dotdash lines. The 50 percent completeness limits, determined by the fitting relationships, are marked with dotted lines, and their values are given in Table~\ref{tab:properties}.}
\label{fig:rgb_completeness}
\end{figure}

To assess the limiting magnitudes and photometric uncertainties for each image channel, a completeness test was conducted. We utilized the photutils \citep{bradley2024} package to add artificial stars to each image. Within the instrumental magnitude range of -14 to -7, 2000 Gaussian-profile artificial stars were added to each image with randomly assigned coordinates, with approximately 1200 artificial stars falling within the image field of view. Subsequently, the artificial stars were recovered using SExtractor \citep{1996A&AS..117..393B} by photometry. This process was repeated 100 times. The limiting magnitudes ($m_{\rm lim}$; defined as the 50 percent completeness level) were derived from all of the artificial star data for each individual image, and the data were fitted to the following interpolation function \citep{1995AJ....109.1044F}:
\begin{equation}
f(m) = \frac{1}{2}\left[1-\frac{\alpha(m - m_{\rm lim})}{\sqrt{1+\alpha^2(m - m_{\rm lim})^2}}\right]
\end{equation}
where $m$ is the photometric magnitude of one of the R, G, and B channels, and $\alpha$ is a parameter that measures how steeply $f(m)$ declines from 1.0 to 0.0. The detected artificial stars were binned into steps of 0.2 mag and then the fraction $f(m)$ of recoverable stars in each bin was calculated. The fitting results of the completeness curves for the three channels are shown in Fig.~\ref{fig:rgb_completeness}, and the limiting magnitudes are given in Table~\ref{tab:properties}. Additionally, the differences in position and magnitude between the initial and recovered artificial stars were evaluated, and the mean and standard deviation of the positional offsets $dx$, $dy$, and magnitude offsets $dm$ for the artificial stars were calculated. The mean positional offsets for the three channels were all zero, and the mean magnitude offsets were all 0.04 mag. The standard deviations for the positional offsets were approximately 0.2 pixels, and the standard deviations for the magnitude offsets were 0.14 mag, with the corresponding values given in Table~\ref{tab:properties}. The instrumental zero points ($m_0$) in the table are provided by subsequent calculations of the extinction coefficients. The photometric magnitudes and limiting magnitudes have been adjusted for the instrumental zero point.

\begin{table}
\caption{Properties of images captured by the cloud monitor.}
\label{tab:properties}
\begin{tabular}{lrrr}
\hline
Property & R & G & B\\
\hline
$m_0$ (mag) 				&  16.34 & 16.73 & 15.91\\
$\alpha$ 				&   1.50 &  1.67 &  1.65\\
$m_{\rm lim}$ (mag)		&   6.98 &  7.13 &  6.57 \\
$\sigma_{dx}$ (pixel)	&   0.21 &  0.19 &  0.22\\
$\sigma_{dy}$ (pixel)	&   0.21 &  0.19 &  0.22\\
$\sigma_{dm}$ (mag)		&   0.14 &  0.14 &  0.14 \\
\hline
\end{tabular}
\end{table}

\section{THE METHOD TO CALIBRATE ALL-SKY IMAGE} \label{sec:method}
\subsection{Models of all-sky camera} \label{sec:model}
The camera imaging model is based on the principles of pinhole imaging, establishing a connection between the real-world coordinates and the image coordinates. This process involves handling transformations across various coordinate systems. The coordinates $(X,Y,Z)$ of points in the WCS are transformed into the camera coordinate system $(x',y',z')$ through rigid body transformations, which include rotation ($\boldsymbol R$) and translation ($\boldsymbol t$). Subsequently, they are transformed into the image coordinate system's coordinates $(x,y)$ through perspective projection. Ultimately, an affine transformation is applied to map these coordinates onto the pixel coordinates $(u,v)$ of the image \citep{888718}. 

The distortion-free projective transformation of a point $\boldsymbol P$ from WCS to a pixel $\boldsymbol p$ within the image plane can be represented by the pinhole camera model (PCM), as follows \citep{888718}:

\begin{equation}
s \boldsymbol {p =  M[R|t]P} 
\end{equation}
or 
\begin{equation}
s \begin{bmatrix}
u \\
v \\
1
\end{bmatrix} =  
\begin{bmatrix}
f_x & 0 & c_x \\
0 & f_y & c_y \\
0 & 0 & 1
\end{bmatrix}
\begin{bmatrix}
r_{11} & r_{12} & r_{13} & t_1 \\
r_{21} & r_{22} & r_{23} & t_2 \\
r_{31} & r_{32} & r_{33} & t_3 \\
\end{bmatrix}
\begin{bmatrix}
X \\
Y \\
Z \\
1
\end{bmatrix}
\end{equation}
where $\boldsymbol M$ is the camera intrinsic matrix, which is composed of the focal lengths 
$f_x$, $f_y$, and the principal point $(c_x,c_y)$, and $s$ is the projective transformation's arbitrary scaling. The joint rotation-translation matrix, denoted as $\boldsymbol{[R|t]}$, comprises the extrinsic parameters $r_{ij}$ for rotation and $t_i$ for translation.

In the process, the perspective projection model of pinhole camera assumes that the rays of light pass through the lens and project onto the image sensor, and the mapping between points $(x',y',z')$ in the camera coordinate system and the normalized image coordinates $(x,y)$ can be expressed as follows:

\begin{equation}
\begin{cases}
\theta \,  =  \,\arctan(\sqrt{x'^2+y'^2}, z')\\
\phi\,  = \, \arctan(y',x') \\
  r\,  = \, f \tan(\theta) \\
  x\,  = \, r\cos\phi \\
  y\,  = \, r\sin\phi
\end{cases}
\end{equation}
where $(\phi, \theta)$ represents the spherical coordinates, $r$ is the radial distance from the optical center, $f$ is the focal length of the imaging system. This model is suitable for most narrow-angle and even wide-angle lenses, but it is not appropriate for fisheye lenses that cover an entire hemisphere. Fisheye lenses are designed to adhere to distinct projection models that effectively translate the hemispherical FOV onto a finite image plane. The radial symmetric projection for these specialized lenses, as detailed in \citet{1642666}, is listed in Table~\ref{tab:projection}. A general projection model for describing these radial projections is presented as follows:
\begin{equation}
r(\theta) =  k_1\theta +k_2\theta^3 +k_3\theta^5 +k_4\theta^7 +k_5\theta^9 +..., \label{eqa:r_theta}
\end{equation}
where $k_i$ are coefficients of radial projection model.

\begin{table}
\caption{The zenithal perspective projections.}
\label{tab:projection}
\begin{tabular}{llc}
\hline
Name & Formula$^1$ & AIPS convention$^2$ \\
\hline
Gnomonic projection      & $r = f \tan(\theta)$ & TAN  \\
Stereographic projection & $r = 2 f \tan(\theta/2)$ & STG  \\
Equidistant projection   & $r = f \theta$ & ARC \\
Equisolid angle projection   & $r = 2 f \sin(\theta/2)$ & ZEA  \\
Orthographic projection  & $r = f \sin(\theta)$ & SIN  \\
\hline
\end{tabular}\\
\scriptsize
1. The projection formulas are provided in \citet{1642666}.\\
2. The corresponding projection abbreviations are used in the AIPS conventions \citep{2002A&A...395.1077C}.
\end{table}

Furthermore, it is essential to account for camera distortion, which is the change in an image's perspective caused by elements such as the camera lens and sensor. The two primary types of distortion are radial and tangential. Radial distortion is characterized by the bending of light rays that pass through the outer regions of the lens to a greater extent than those through the center. Tangential distortion, also known as decentering distortion, emerges from an imperfect alignment between the lens assembly and the image plane. This misalignment leads to a non-radial image distortion, manifesting as rotation and skewing effects within the image. To correct for these distortions, several polynomial models have been developed and are widely utilized \citep{Brown1971,1642666}.

\subsection{AIPS convention}
\citet{2002A&A...395.1077C} provides a method for the conversion between pixel coordinates $(u,v)$ and celestial coordinates in the AIPS convention. This conversion process involves a transformation from pixel coordinates to projection plane coordinates $(x,y)$, which parallels the process used in the PCM. The transition is represented as follows: 

\begin{equation}
\begin{bmatrix}
x \\
y
\end{bmatrix}  = 
\begin{bmatrix}
\rm CD1\_1 & \rm CD1\_2 \\
\rm CD2\_1 & \rm CD2\_2
\end{bmatrix}
\begin{bmatrix}
u + f(u,v) \\
v + g(u,v) \label{eqa:uv_fg}
\end{bmatrix}
\end{equation}
where CD$i\_j$ are elements of a linear transformation matrix with scale. The functions $f(u,v)$ and $g(u,v)$ encapsulate the quadratic and higher-order terms of the distortion polynomial \citep{2005ASPC..347..491S}, and are expressed as:

\begin{equation}
\begin{cases}
  f(u,v)\,  =  \,
  \sum_{p,q} {\rm A}_{p,q}u^pv^q, \qquad 
  p+q\leq {\rm A}_{\rm ORDER}  \\
  g(u,v) \, =  \,
  \sum_{p,q} {\rm B}_{p,q}u^pv^q, \qquad
  p+q\leq {\rm B}_{\rm ORDER} \label{eqa:uv_pq}
\end{cases}
\end{equation}
where ${\rm A}_{p,q}$ and ${\rm B}_{p,q}$ are the polynomial coefficients for polynomial terms $u^pv^q$. The projection coordinates $(x,y)$ are then subjected to a spherical projection transformation (as listed in Table~\ref{tab:projection}) to yield native spherical coordinates $(\phi,\theta)$. These coordinates are subsequently rotated to obtain the celestial coordinates. 

Contrasting with the PCM, the AIPS model employs a distinct sequence in the application of affine transformations and distortions. It applies the distortion terms directly to the pixel coordinates $(u,v)$, thereby facilitating the conversion. For an all-sky camera equipped with a fisheye lens, an appropriate projection model within the AIPS convention can be selected to suit the specific requirements.

\subsection{Method}
\begin{figure*}
\centering
\includegraphics[width=0.9\textwidth]{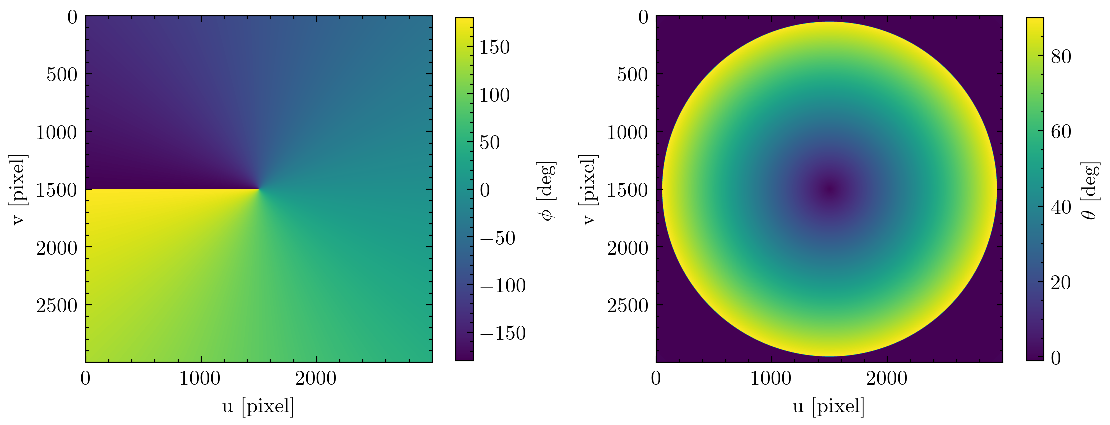}
\caption{The image is represented in native spherical coordinates.}
\label{fig:img_polar}
\end{figure*}

A straightforward thought is to segment the all-sky image and then utilize Astrometry.net \citep{2010AJ....139.1782L} to identify stars in each segment. This approach would yield a star catalog with a one-to-one correspondence between the image coordinates and celestial coordinates, which can be used for the calibration of the all-sky image. The challenge with this idea lies in the fact that Astrometry.net can only discern stars within a limited area around the projection center. Within 30 degrees of the zenith, the radial variations of different projections are very similar, but beyond 30 degrees, there are noticeable differences between projections (as shown in Figure 1(a) of \citealt{1642666}). Stars in the zenith area can be identified by Astrometry.net, but segments beyond the zenith cannot be directly used for star identification due to the deviation from the projection center. These segments require transformation to meet the requirements for Astrometry.net to identify stars. This method leverages the HEALPix \citep{2005ApJ...622..759G} projection to map the all-sky image onto a spherical surface, and by performing uniform segmentation and image transformation, it realizes the idea of all-sky image calibration. The detailed process is as follows: 

To begin with, it is essential to identify the projection center and the associated projection transformation for the all-sky image. Since the center of FOV remains unchanged, it only needs to be determined once. The projection center of the all-sky image is typically in close proximity to the center of the FOV $(u_0,v_0)$, which can be approximated from the field's center, and this point is also near the zenith position. The projection transformation is determined by the fisheye lens, with common projections shown in Table \ref{tab:projection}. If using a general form of radial projection (Equation~\ref{eqa:r_theta}), a number of aligned stars must be selected to obtain their image coordinates and the corresponding zenith angles (associated with the observation site and time of the image), and then fit and solve the coefficients of Equation~\ref{eqa:r_theta} using the least squares method. In this study, the Sigma fisheye lens used in the all-sky camera employs an equisolid angle projection, which is represented by $2f\sin(\theta/2)$. Consequently, the transformation between the native spherical coordinates $(\phi,\theta)$ and the pixel coordinates can be expressed as:
 
\begin{equation}
\begin{cases}
  u\,  = \, u_0 + r\cos\phi \\
  v\,  = \, v_0 + r\sin\phi \\
  r\,  = \, R\sin(\theta/2) \\
\theta \,  =  \,2\arcsin(r/R)\\
\phi\,  = \, \arctan(v-v_0,u-u_0) \label{eqa:10}
\end{cases}
\end{equation}
where $R$ is the conversion coefficient from projection coordinates to pixel coordinates, and $r$ is the distance from $(u,v)$ to the projection center, calculated as $r=\sqrt{(u-u_0)^2+(v-v_0)^2}$. The value of $R$ can be estimated based on the pixel radius of the FOV boundaries ($r_{\rm max}$) and the angular size of the FOV ($\theta=\pi/2$). The value of $R$ can also be calculated through the equisolid projection relationship $R=2f/d_{\rm px}=2098$ px \citep{2019A&A...626A.105B}, with the camera focal length $f=4.5mm$ and the detector's pixel size being $d_{\rm px}=4.29\mu m$. The native spherical coordinates corresponding to the pixels in the all-sky image are displayed in Fig.~\ref{fig:img_polar}.

\begin{figure*}
\centering
\includegraphics[width=0.8\textwidth]{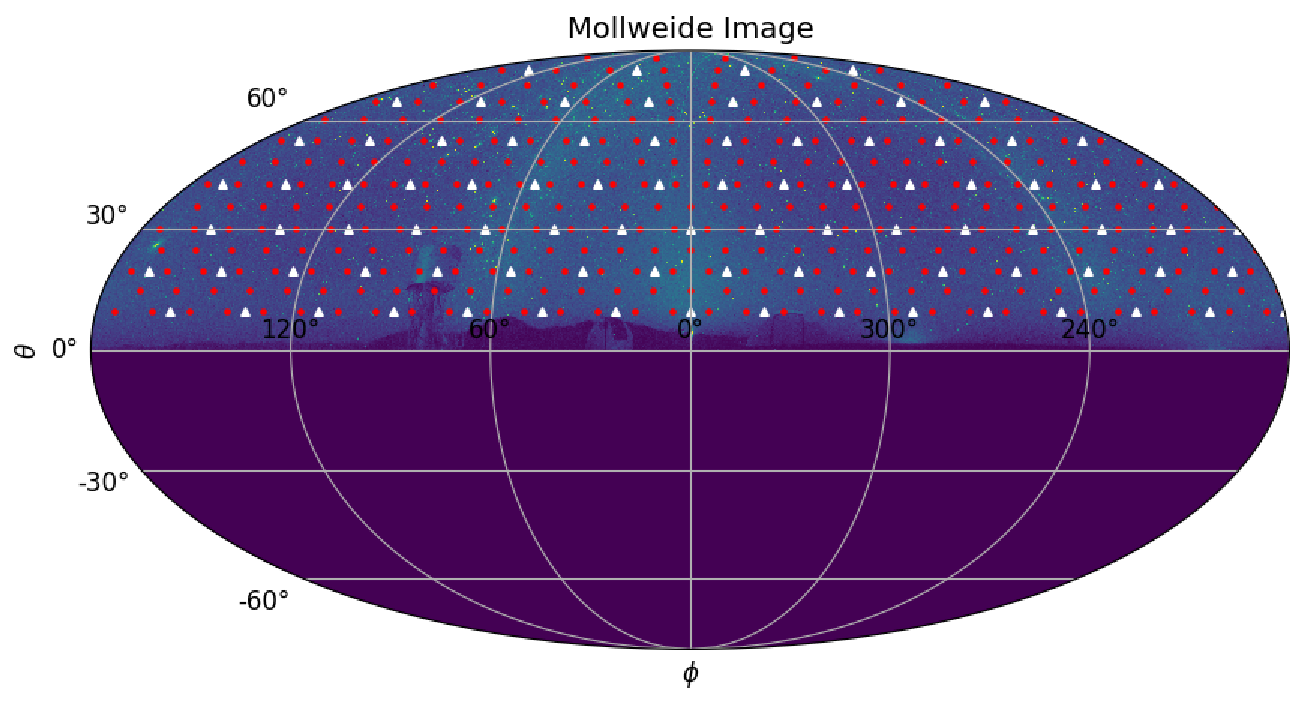}
\caption{The clear-sky image is displayed using the Mollweide projection. White triangles and red dots represent the pixel layout of HEALPix at resolutions of nside four and eight, respectively.}
\label{fig:heal_img}
\end{figure*}

Next, we employ HEALPix to interpolate the image, enabling us to segment the zenith region post-rotation. We construct a HEALPix image with an nside parameter set to 1024, utilizing angular coordinates. The angular coordinates can be used as the spherical coordinates $(\phi_{\rm H},\theta_{\rm H})$ in Equation~\ref{eqa:10}, allowing for the calculation of the corresponding pixel coordinates $(u_{\rm H},v_{\rm H})$. The HEALPix image can be derived from the all-sky image using bilinear interpolation techniques, as shown in Fig.~\ref{fig:heal_img}. Additionally, the HEALPix maps of $u_{\rm H}$ and $v_{\rm H}$ are obtained. The pixel resolution of the HEALPix image is $0\fdg057$\,pixel$^{-1}$, which is slightly higher than the resolution of the all-sky image at $0\fdg064$ per pixel. 

In the third step, we rotate different positions of the HEALPix image to the zenith for bright star identification and coordinate solving. Simultaneously, the HEALPix maps of $u_{\rm H}$ and $v_{\rm H}$ also undergo the same rotation transformation, resulting in the same rotated coordinates $(u_{\rm rot}, v_{\rm rot})$ as the HEALPix image. Once we have obtained the rotated coordinates of the stars, we can perform interpolation on $u_{\rm H}$ and $v_{\rm H}$ separately to derive the pixel coordinates $u$ and $v$ of the stars. The red dots of Fig.~\ref{fig:heal_img} represent the pixel positions of the HEALPix image when the nside is eight, which also defines the solution area centered on the red dot, encompassing over 300 areas. Each area is sequentially rotated to the zenith and projected onto a plane using the gnomonic projection, with Astrometry.net employed to identify the stars within the projected area, and the area with a sparse number of stars will be skipped. We have adopted the 4100-series index built from the Tycho-2 star catalog \citep{2000A&A...355L..27H} as the pre-indexed reference, which is suitable for images wider than one degree. To improve the computational speed, the settings here are as follows: the area size is set to 600 by 600\,pixels, corresponding to an approximately $10^{\circ} \times10^{\circ}$ region on the celestial sphere, with a pixel scale ranging from 75 to 105\,arcseconds, and a time limit of 20\,seconds is imposed.

\begin{figure}
\centering
\includegraphics[width=0.46\textwidth]{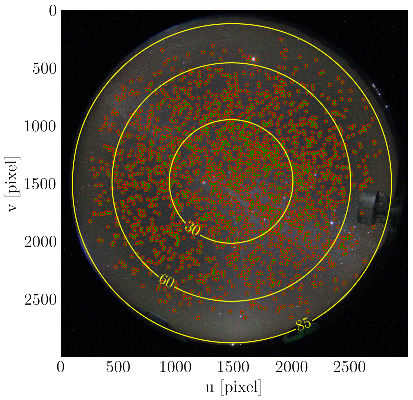}
\caption{The distribution of stars used for coordinate calibration across the all-sky image. Red circles denote the top 2100 stars by flux identified by Astrometry.net, and green plus signs represent the positions calculated by the calibration model. The yellow circles, from the center outwards, represent the zenith angles of 30, 60, and 85\,degrees respectively.}
\label{fig:img_source}
\end{figure}

Afterward, we merge the star catalogs derived from each region, eliminating any duplicate stars to compile a comprehensive catalog for the all-sky image. This star catalog encompasses parameters such as the pixel coordinates, equatorial coordinates derived from the Tycho-2 star catalog, stellar flux, and matching weights.  Based on the geographical coordinates of the observatory and the observation time of the image, the horizontal coordinates of the stars are derived from the equatorial coordinates and added to the star catalog. With the total number of identified stars surpassing 5000, we select the top 2100 stars (T2100S) based on their flux for the purpose of coordinate calibration, as indicated by the red circles in Fig.~\ref{fig:img_source}. The density of stars diminishes progressively from the zenith outward, with a notable scarcity beyond the 80$^{\circ}$ zenith angle.

\begin{figure*}
\centering
\includegraphics[width=0.9\textwidth]{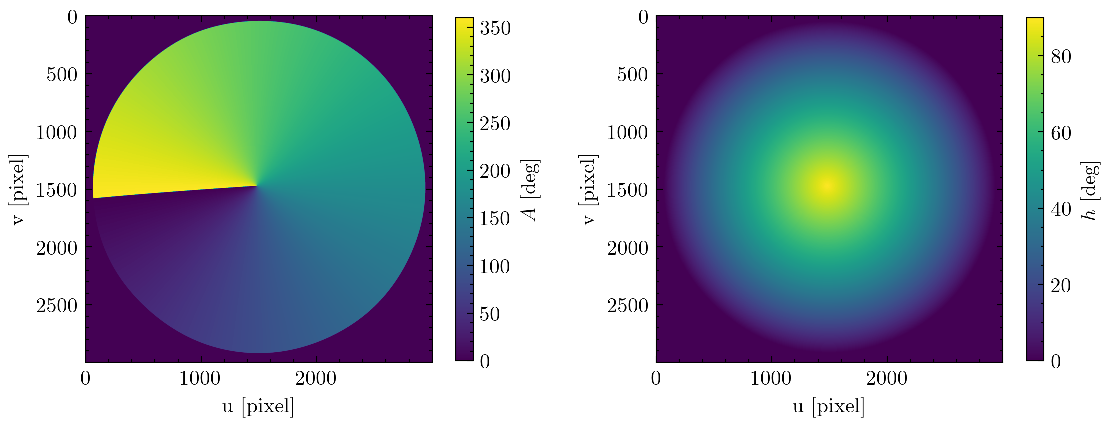}
\caption{The horizontal coordinates $(A,h)$ obtained through the transformation of GWCS.}
\label{fig:img_coordinates}
\end{figure*}

Finally, we employed the Generalized World Coordinate System (GWCS; \citealt{dencheva2024}) to calibrate the all-sky image. The GWCS implements the AIPS model and facilitates direct calibration using the pixel coordinates and horizontal coordinates of stars. The calibration coordinates of stars are indicated by the green plus signs in Fig.~\ref{fig:img_source}, with positional deviations of less than one pixel, and the corresponding error analysis will be discussed in the following section. Using the transformation of GWCS, we calculated the horizontal coordinates $(A,h)$ for the all-sky image, as shown in Fig.~\ref{fig:img_coordinates}.

\subsection{Calibration deviations}
During the calibration process of an individual all-sky image, the positional deviations of stars primarily stem from two sources. The first is the transformations involved during star selection, including image projection, rotation, and interpolation. The second source is the fitting deviation of the calibration software tools.

\begin{figure}
\centering
\includegraphics[width=0.46\textwidth]{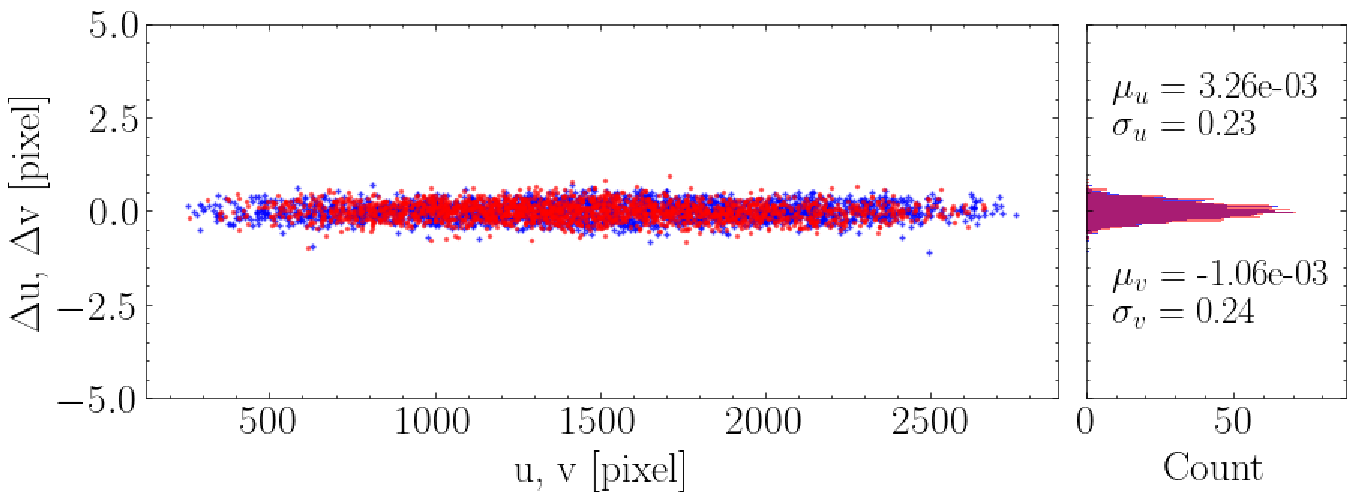}
\includegraphics[width=0.46\textwidth]{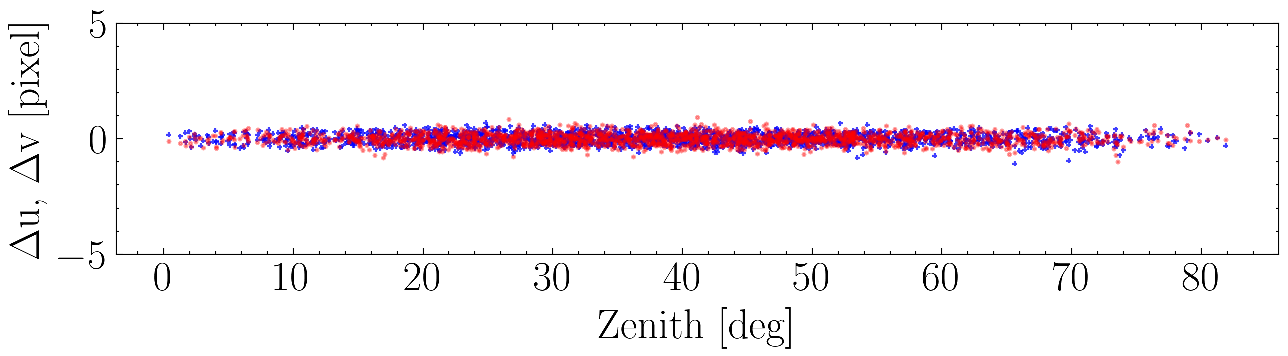}
\includegraphics[width=0.46\textwidth]{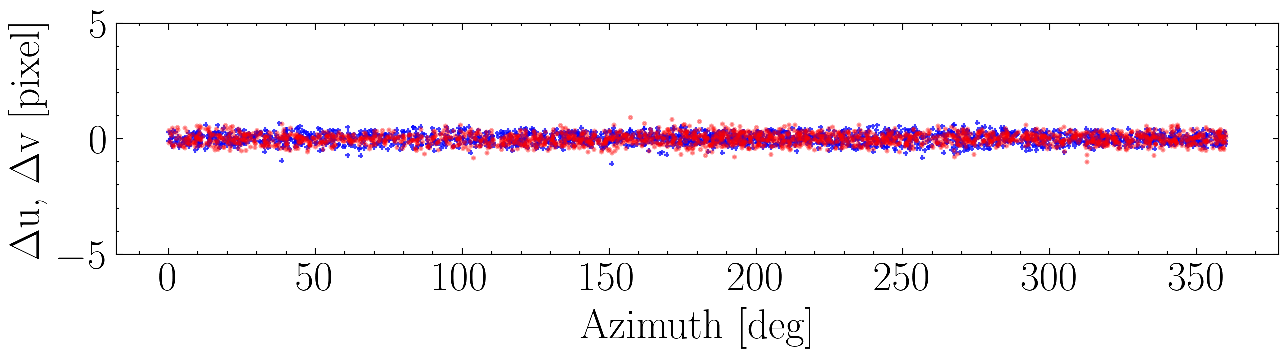}
\caption{The positional deviations of stars caused by image transformations vary with the stellar image position, zenith angle, and azimuth angle, along with their statistical distribution.}
\label{fig:pixdist}
\end{figure}

We employed SExtractor for photometry on the all-sky image to ascertain the initial coordinates $(u, v)$ of the stars within the image when calculating the deviations attributable to image transformations. For the stars after transformation, we adopted the merged star catalog's T2100S, utilizing their transformed coordinates $(u', v')$ along with the corresponding zenith and azimuth angles. We then compared the positional discrepancies $(\Delta u, \Delta v)$ of these stars. The deviations in star positions due to image transformation, as they relate to the coordinates $(u, v)$, zenith, and azimuth angles, are illustrated in Fig.~\ref{fig:pixdist}. The mean positional deviations of the stars attributable to image transformation were close to zero pixels, with standard deviations of 0.23 and 0.24\,pixels, respectively. The positional deviations showed no clear variation with zenith and azimuth angles, suggesting that the image transformation is not influenced by the differential zenith and azimuth angles of the stars.

\begin{figure*}
\centering
\includegraphics[width=0.52\textwidth]{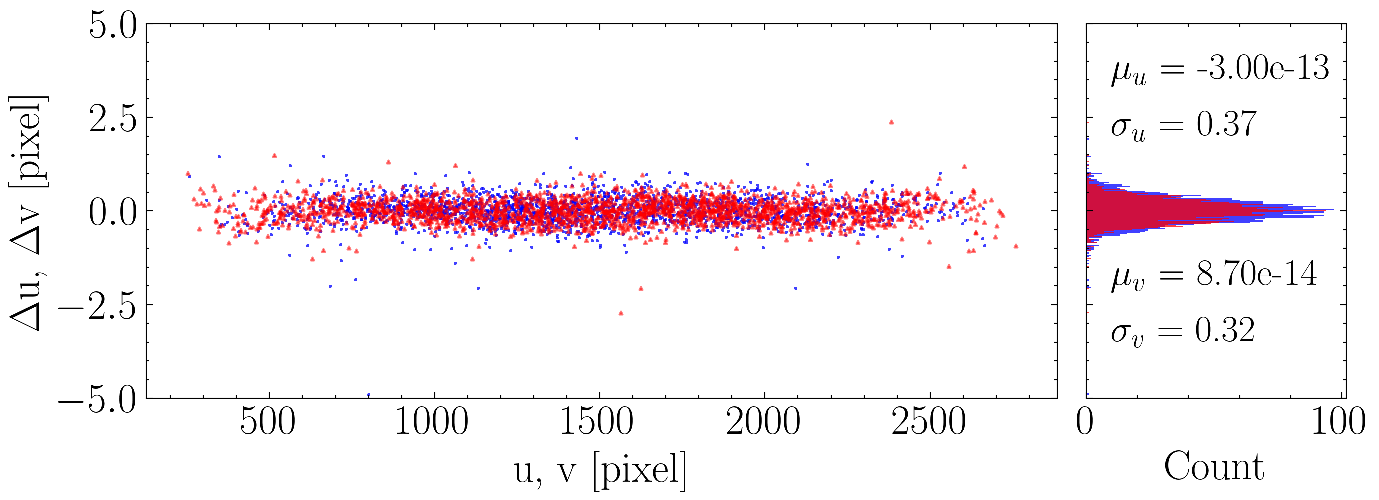}
$\quad$
\includegraphics[width=0.44\textwidth]{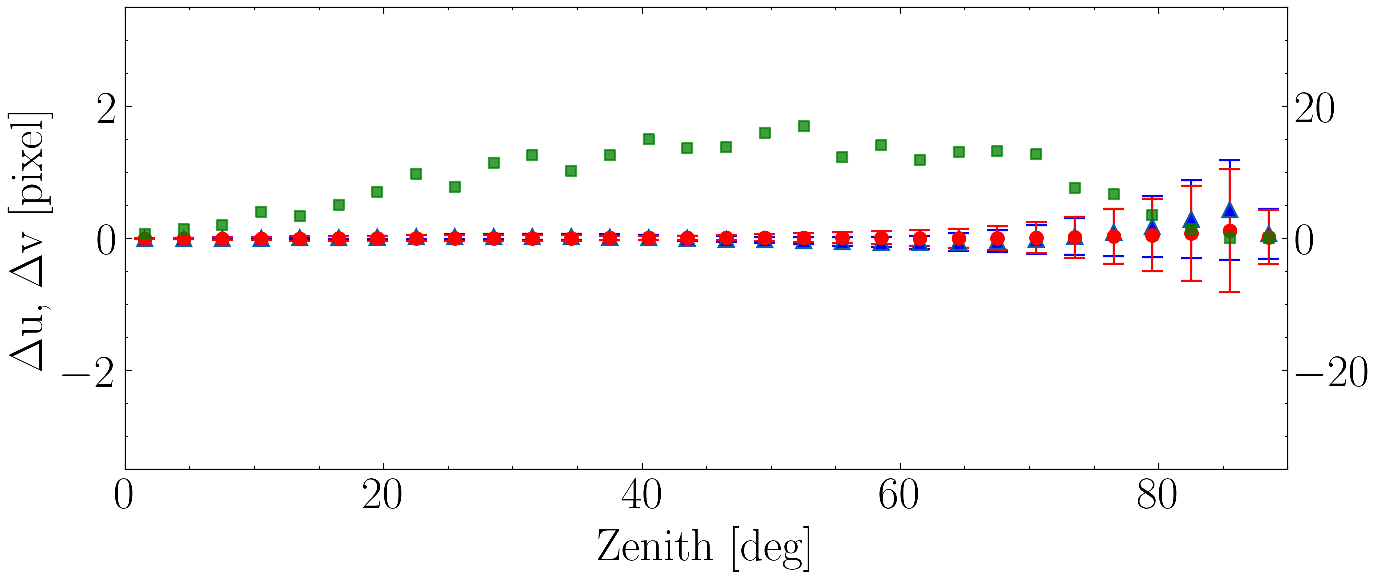}
\caption{The left panel displays the calibration deviations of the all-sky image in the u and v directions, accompanied by their statistical distributions. The right panel illustrates the positional deviations plotted against the zenith angle. The red circles and blue triangles represent the mean deviations in the u and v directions, respectively, with the error bars indicating the corresponding standard deviations. The green squares denote the star density, measured per 100 square degrees, and are aligned with the scale on the secondary y-axis.}
\label{fig:pixgwcs}
\end{figure*}

To assess the fitting deviations, we applied the inverse transformation of the GWCS fitting tool to the horizontal coordinates $(A,h)$ of the T2100S stars in the merged star catalog, deriving their pixel coordinates, which were then compared with the corresponding coordinates $(u', v')$ in the merged star catalog. The comparative analysis is depicted in the left panel of Figure \ref{fig:pixgwcs}, which indicates that GWCS  achieves a mean deviation of zero and a standard deviation of approximately 0.3\,pixels for both u and v axes. The positional deviation is primarily manifested at high zenith angles and is shown to vary with the zenith angle in the right panel of Fig.~\ref{fig:pixgwcs}. In the figure, the zenith angle's annular intervals are separated by 3$^{\circ}$, and the error bars represent the standard deviation of the positional deviations. The mean positional deviation remains close to zero across various zenith angles, while the standard deviation is less than 0.25\,pixels within a 70$^{\circ}$ zenith angle. Beyond the 70$^{\circ}$ zenith angle, there is an increase in the standard deviation, predominantly due to the reduction in the number of stars, which does not offer adequate constraints for the calibration model in areas beyond this threshold. The green squares in the diagram represent the stellar density for each annular region, scaled according to the right-hand side vertical axis. The star densities at zenith angles of 70 and 80\,degrees are 13 and three stars per 100 square degrees, respectively.  

\begin{figure*}
\centering
\includegraphics[width=0.52\textwidth]{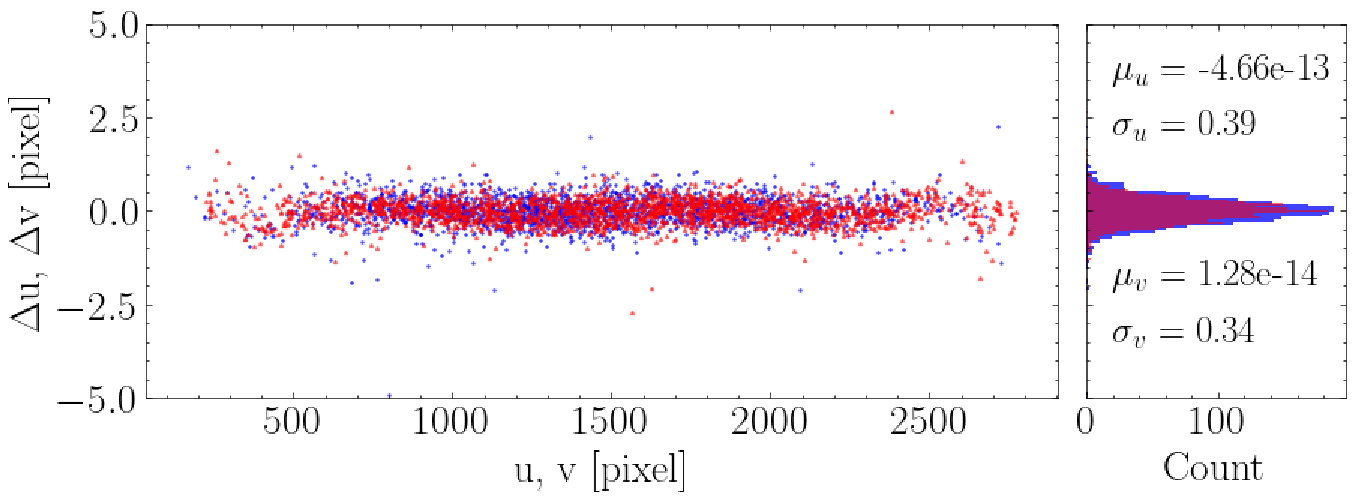}
$\quad$
\includegraphics[width=0.44\textwidth]{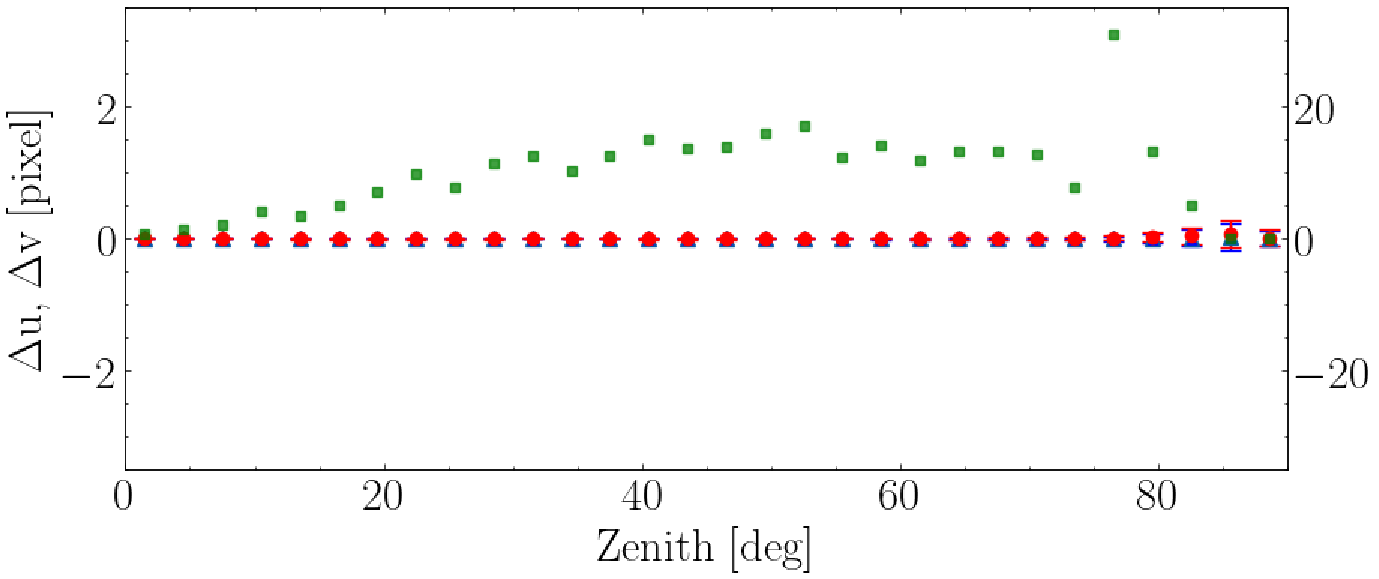}
\caption{Consistent with Figure \ref{fig:pixgwcs}. The image calibration includes an additional four images of stars with zenith angles greater than 75 degrees.}
\label{fig:npix_img_gwcs}
\end{figure*}

We attempted to include additional stars near the zenith to constrain the model and improve the fitting deviations there. The stars were selected from the calibration star catalog of four images taken at nearby times, with star flux ranking within the top two thousand and zenith angles greater than 75 degrees. A total of 86 stars were added. Following the same process to calculate the fitting deviations, the results, as shown in Fig.~\ref{fig:npix_img_gwcs}, indicate that the added stars significantly improved the fitting deviations when the zenith angle was greater than 75 degrees. The standard deviation of the fitting deviations at a zenith angle of 85 degrees was reduced from 0.93 pixels to 0.21 pixels.

\subsection{Selection of star identification areas}
The method employs bright stars from an all-sky alignment for model fitting, but the process is time-consuming due to the numerous segmented areas used for image transformation and star identification. For two-dimensional polynomial fitting, the minimum number of fitting points needed depends on the degree of the polynomial. With a 6th-degree polynomial used for model fitting, over 40 matching star points are required (Equations~\ref{eqa:uv_fg} and \ref{eqa:uv_pq}). Therefore, we aim to achieve time-saving by minimizing the star identification areas while ensuring a reasonable fitting deviation.

We tested the impact of segment selection by setting the nside value in HEALPix, adding extra equally spaced segments, and rotating the segment regions. The assessment was performed by calculating the mean squared difference (MSD) between the photometric positions and the fitted positions of the top 2000 stars. The assessment outcomes are presented in Table \ref{tab:selection}, which includes the values of nside, the number of segmentation regions, the number of stars identified in the segmentation regions, the evaluated MSD, the number of stars when the segmentation regions are incremented by ten degrees in the azimuthal direction, and the corresponding MSD. When nside is set to eight, it corresponds to a complete all-sky star identification with overlaps in segmentation regions, yielding an MSD of 0.266, which is also the lowest value. At nside seven, the segmentation regions nearly cover the entire sky, with the MSD also registering at 0.266. Due to the non-uniform distribution of stars across the sky, this may impact the fitting results. Comparing the MSD with the MSD when the azimuth is increased by ten degrees, the impact is minimal for nside values greater than three. For nside two, the effect of the non-uniform star distribution is more pronounced. Adding 12 segmentation regions with a 15-degree elevation and equal spacing for nside two, three, and four scenarios results in improved MSD, with the most significant improvement observed for nside two. Using 1.1 times the lowest MSD as a threshold, star identification using nside = 3 with additional segmentation regions or nside = 4 can enhance the efficiency of the method. 

\begin{table}
\caption{Star selection for model fitting.}
\label{tab:selection}
\begin{tabular}{crrcrc}
\hline
nside & N$_{\rm region}$ & N$_{\rm star}$ & MSD & N$_{\rm star}$(+10$^{\circ}$) & MSD (+10$^{\circ}$)\\
\hline
2 &  20 &  205 & 0.652 &  188 & 0.932 \\
2 &  20+12 &  220 & 0.405 &  206 & 0.600 \\
3 &  48 &  579 & 0.294 &  541 & 0.319 \\
3 &  48+12 &  594 & 0.289 &  559 & 0.297 \\
4 &  88 & 1143 & 0.273 & 1112 & 0.285 \\
4 &  88+12 & 1148 & 0.272 & 1121 & 0.283 \\
5 & 140 & 1603 & 0.268 & 1591 & 0.272 \\
6 & 204 & 1895 & 0.266 & 1881 & 0.272 \\
7 & 280 & 2017 & 0.266 & 2018 & 0.266 \\
8 & 336 & 2060 & 0.266 & 2060 & 0.266 \\
\hline
\end{tabular}
\end{table}

\section{APPLICATIONS} \label{sec:application}
The method of automatic calibration of all-sky images facilitates image registration and is also the foundation of all-sky image analysis. It can be applied to the study of cloud cover, atmospheric extinction, night sky brightness, transient sources, variable stars, and so on, providing their temporal variations and spatial distribution characteristics. Here, we present the application results for atmospheric extinction, night sky brightness, and known variable stars.

\subsection{Atmospheric extinction}
\begin{figure*}
\centering
\includegraphics[width=0.96\textwidth]{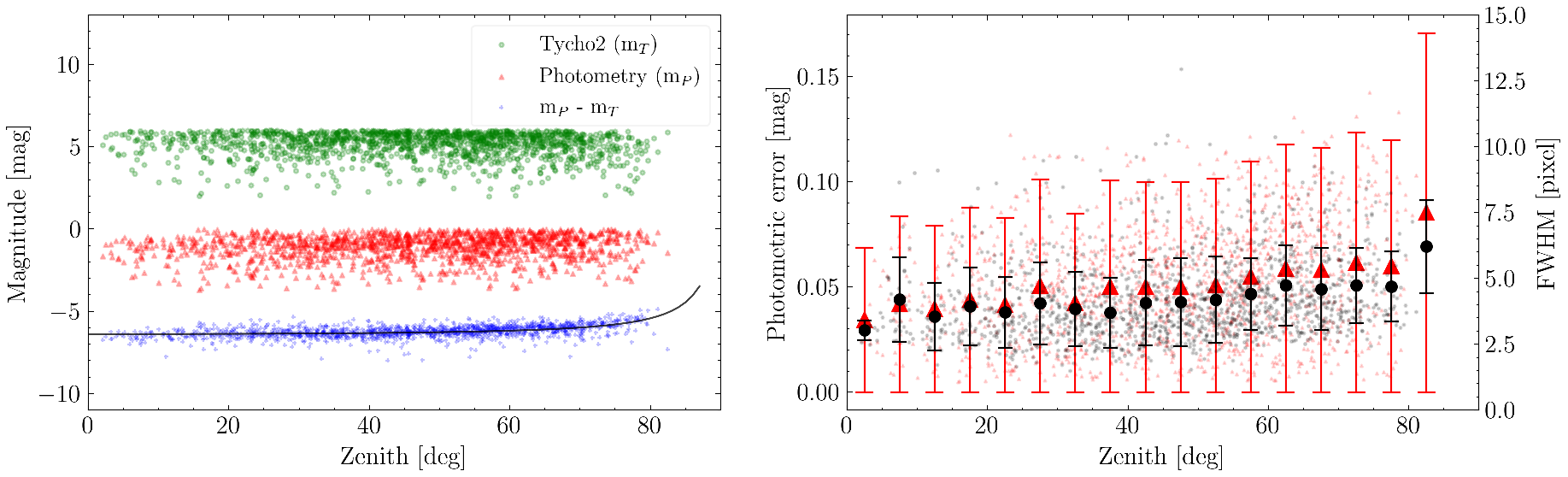}
\caption{Estimation of atmospheric extinction from a single all-sky image. The left panel shows the variation of photometric magnitude with the zenith angle. The green dots represent the star magnitudes from the Tycho2 catalogue ($m_{\rm T}$), the red triangles indicate the photometric magnitudes ($m_{\rm P}$; $m_{\rm P}$ has been increased by a constant of ten for display purposes), the blue plus signs denote the differences between the two magnitudes ($m_{\rm P}-m_{\rm T}$), and the black line represents the fit for atmospheric extinction. The right panel displays the variation of photometric errors and FWHM with the zenith angle. The red triangles with error bars correspond to the photometric errors, while the black dots with error bars represent the FWHM, which is plotted on the right y-axis. The error bars are spaced at intervals of five degrees in zenith angle.}
\label{fig:mag_zen}
\end{figure*}

\begin{figure}
\centering
\includegraphics[width=0.45\textwidth]{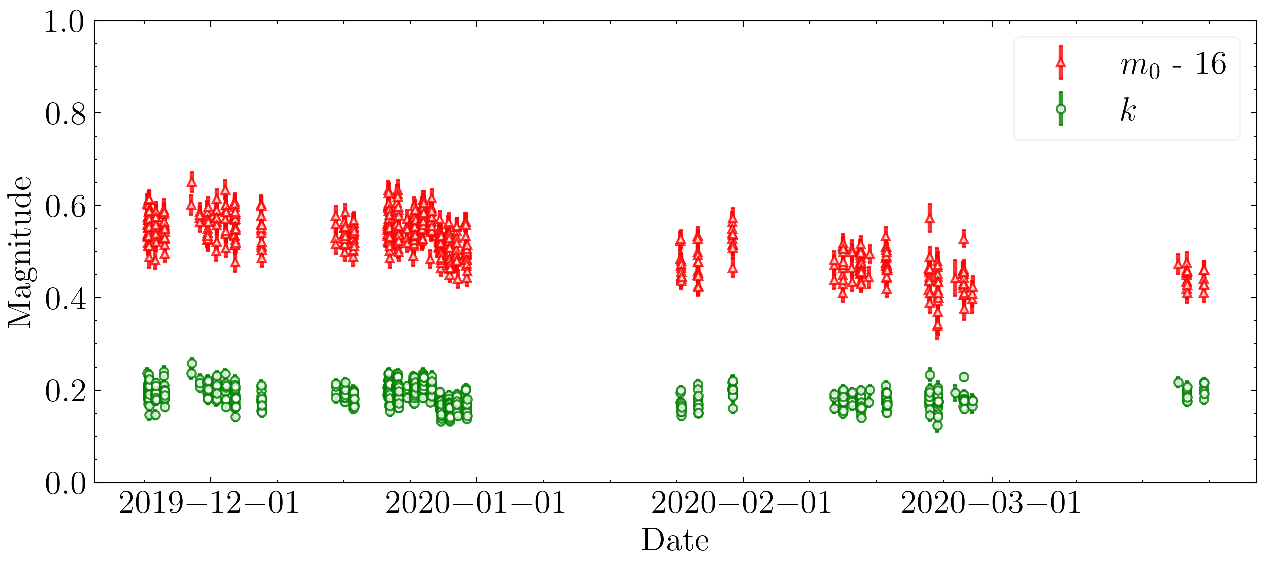}
\caption{Variation of instrumental zero point and atmospheric extinction over time. The instrumental zero point has been offset by a constant of 16 for display purposes.}
\label{fig:extinction}
\end{figure}

The photometry results of the all-sky image are obtained using SExtractor, and they are matched with stars from the Tycho2 catalog that are within a magnitude of six. The matching radius is set to 1$^{\prime\prime}$. The difference between the photometric apparent magnitude $m_{\rm P}$ and the catalog magnitude $m_{\rm T}$ of the star is primarily due to atmospheric extinction, and thus can be expressed as \citep{2011MNRAS.412...33F}:
\begin{equation}
m_{\rm T} - m_{\rm P} = m_0 - k\chi
\end{equation}
where $m_0$ is the instrumental zero point, $k$ is the extinction coefficient, and 
$\chi$ is the airmass. The airmass is represented as a function of the zenith angle 
$\theta$ as \citep{1991cacu.book.....B}:  
\begin{equation}
\chi = \sec \theta(1-0.0012\tan^2\theta)
\end{equation}
The instrumental zero point estimated from the photometry results of a single image is 16.60, and the extinction coefficient is 0.22, as shown in Fig.~\ref{fig:mag_zen}. The figure also presents the photometric errors and the full width at half maximum (FWHM) of the star's point spread function (PSF), both of which increase with the zenith angle. The variations of the instrumental zero point and atmospheric extinction over time are presented in Fig.~\ref{fig:extinction}. The long-term variation in the instrumental zero point may be caused by changes in ambient temperature, while the atmospheric extinction is relatively stable, with a mean of 0.20 and a standard deviation of 0.031.

\subsection{Night sky brightness}
After aligning the instrumental zero point ($m_0$) of the camera, we estimated the night sky brightness for moonless nights. The sky brightness of the all-sky image can be extracted from the background counts per second per arcsec per arcsec$^2$ ($I_{\rm sky}$), and the sky brightness is given by $m_{\rm sky}$ \citep{2011MNRAS.412...33F}:
\begin{equation}
m_{\rm sky} = m_0 -2.5\log(I_{\rm sky})
\end{equation}
The $I_{\rm sky}$ can be calculated by means of the flux of the pixel and its area. The change of solid angle per pixel in the image can be represented as \citep{Feister2000}:
\begin{equation}
\frac{d\Omega}{d{\rm S}} = \frac{\sin\theta}{r'(\theta)r(\theta)}\cdot\frac{\pi}{180}
\end{equation} 
where $r(\theta)$ is the radial projection in Equation \ref{eqa:10}, and $r'(\theta)$ is the derivative of $r(\theta)$. Each pixel covered an area of sky of 38\,683\,arcsec$^2$, which is stable in this research. We used the method of \citet{Skidmore2011} to create a 10th percentile median image from 20 moonless night images, which is the sky background brightness distribution map.

\begin{figure}
\centering
\includegraphics[width=0.45\textwidth]{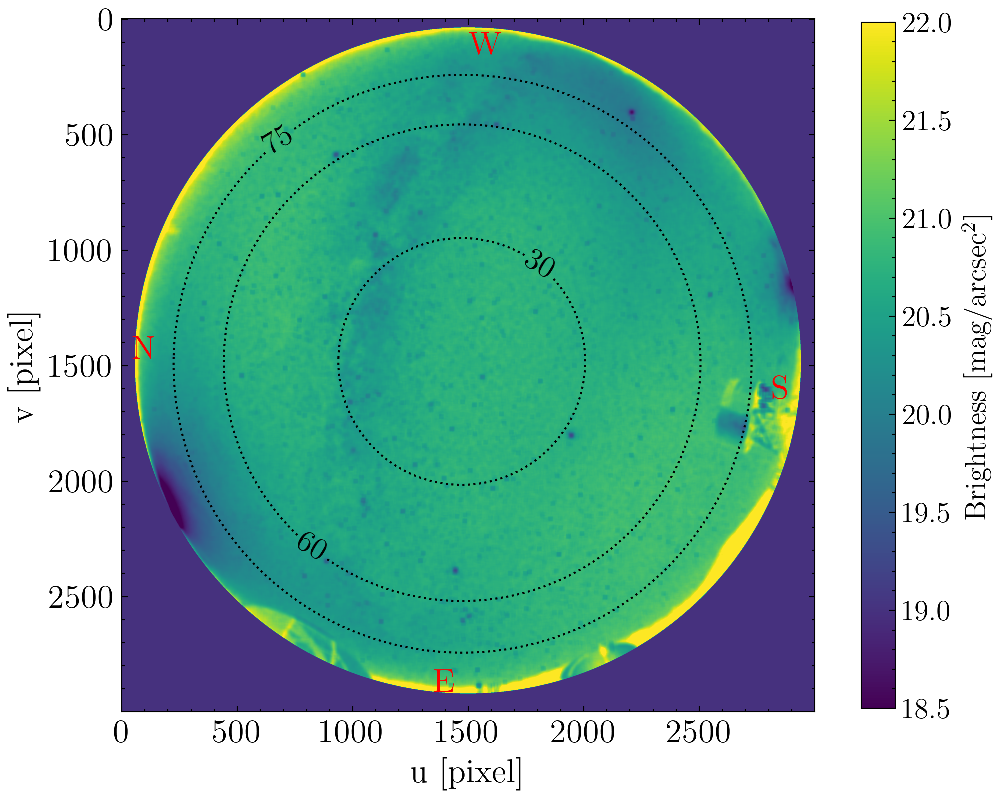}
\includegraphics[width=0.45\textwidth]{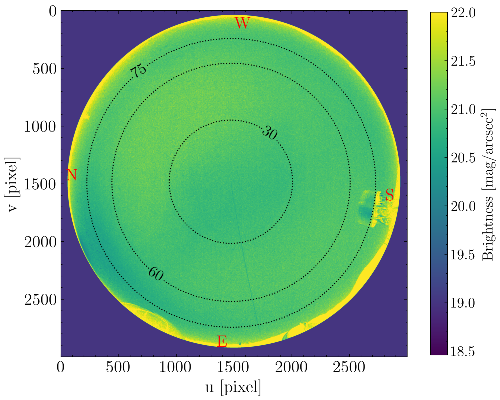}
\caption{The upper and lower figures are the night sky brightness distribution maps of a single image and the merged image from 20 all-sky images, respectively. The three dotted circles represent the 30-degree, 60-degree and 75-degree zenith angles.}
\label{fig:brightness}
\end{figure}

The night sky brightness for a single image and the 10th percentile median image from 20 all-sky images are displayed in Fig.~\ref{fig:brightness}. In addition to the light from stars and the Milky Way, the single image also shows light pollution from the Shiquanhe City in the northeast direction and the temporary construction to the south of the site. From the merged image, the brightness of the zenith area is slightly higher due to the contribution of the Milky Way, with an average sky brightness of 20.94\,mag arcsec$^{-2}$. The brightness in the area between 30 to 60\,degrees from the zenith is slightly dimmer, with an average sky brightness of 21.04\,mag arcsec$^{-2}$. The sky brightness near the 75-degree zenith angle brightens due to atmospheric refraction, with a sky brightness of 20.91\,mag arcsec$^{-2}$. The nadir is mainly composed of mountains and buildings, contributing no brightness, whereas the direction towards Shiquanhe city exhibits significant light pollution, resulting in a sky brightness brighter than 19.5\,mag arcsec$^{-2}$.

\subsection{Known variable stars}
Long-term all-sky monitoring helps to record occasional transient sources, such as gamma-ray bursts, supernova explosions, as well as meteor fireballs and airglow, which are among the light variation events. Naked-eye variable stars are the most common sources of light variation and are detectable by our cloud camera. Here, we present the monitoring results of known variable stars. 

\begin{figure*}
\centering
\includegraphics[width=0.8\textwidth]{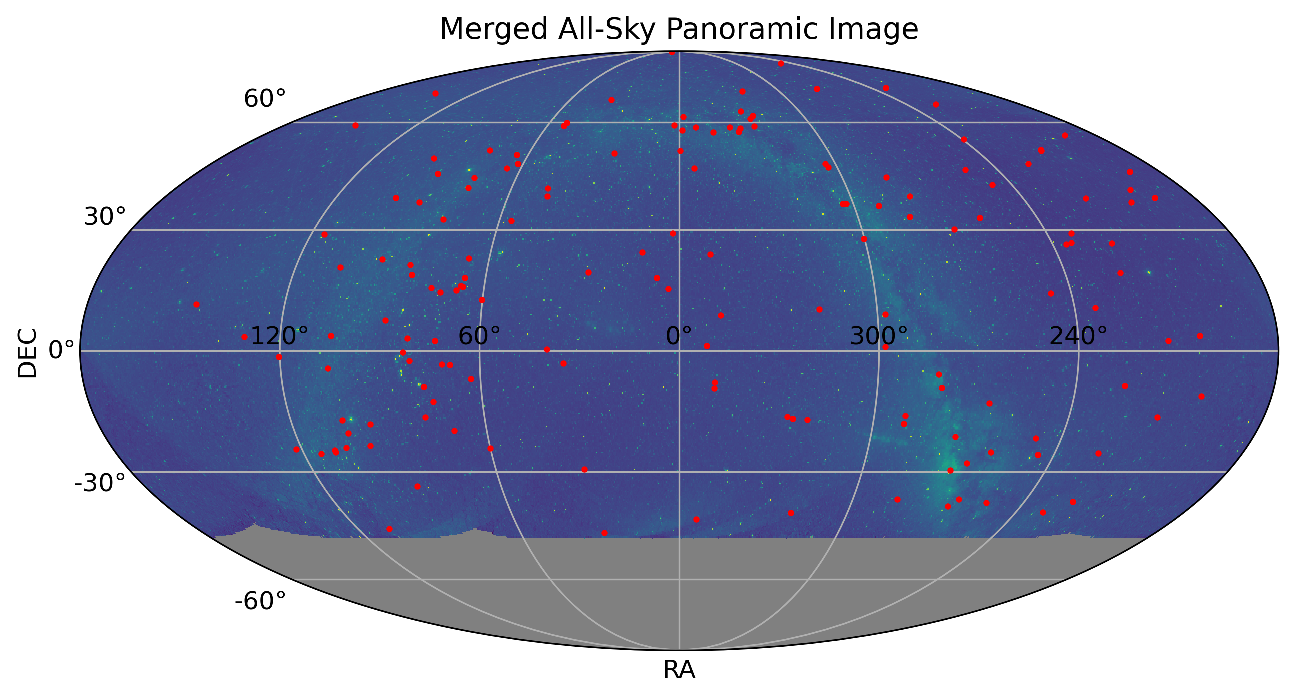}
\caption{The Mollweide projection displays the composite panoramic image, created by merging 38 individual all-sky images. The red dots represent variable stars with brightness greater than five magnitudes.}
\label{fig:heal_var}
\end{figure*}

Observations were conducted from November 2019 to December 2020, selecting over 3,800 all-sky images with star detection counts exceeding 8,000. The star catalogs of these all-sky images were matched with the catalog GCVS \citep{2017ARep...61...80S} with a matching radius of two pixels. Variable stars with a maximum magnitude greater than five and observed more than 100 times are 157, as indicated by the red dots in Fig.~\ref{fig:heal_var}. The background image of Fig.~\ref{fig:heal_var} is a panoramic image obtained by averaging 38 all-sky images, with buildings near the horizon of the all-sky images removed.

\begin{figure*}
\centering
\includegraphics[width=0.7\textwidth]{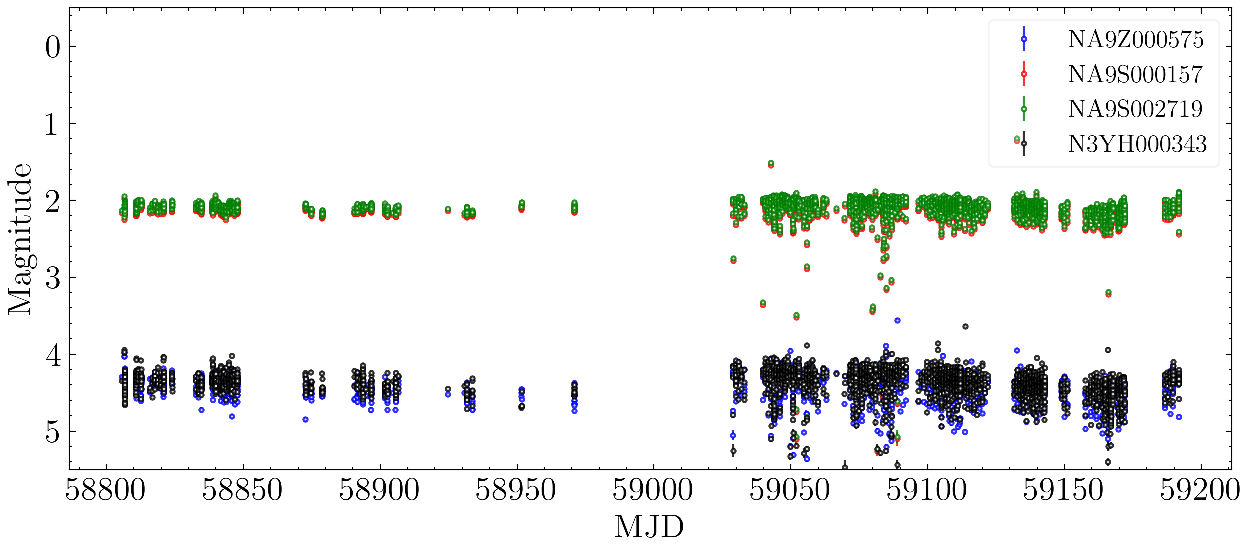}
\caption{The fluctuations in magnitude of the four standard stars around Polaris during the observation period.}
\label{fig:gsc2}
\end{figure*}

\begin{figure*}
\centering
\includegraphics[width=0.45\textwidth]{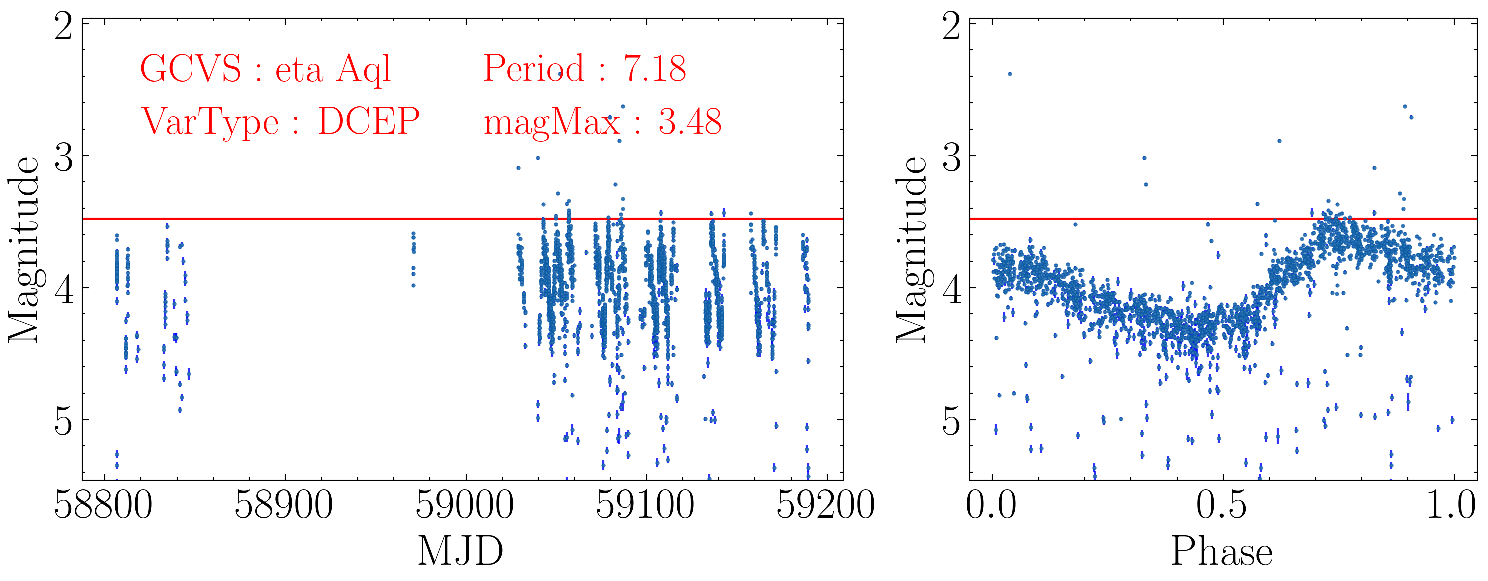}
$\quad$
\includegraphics[width=0.45\textwidth]{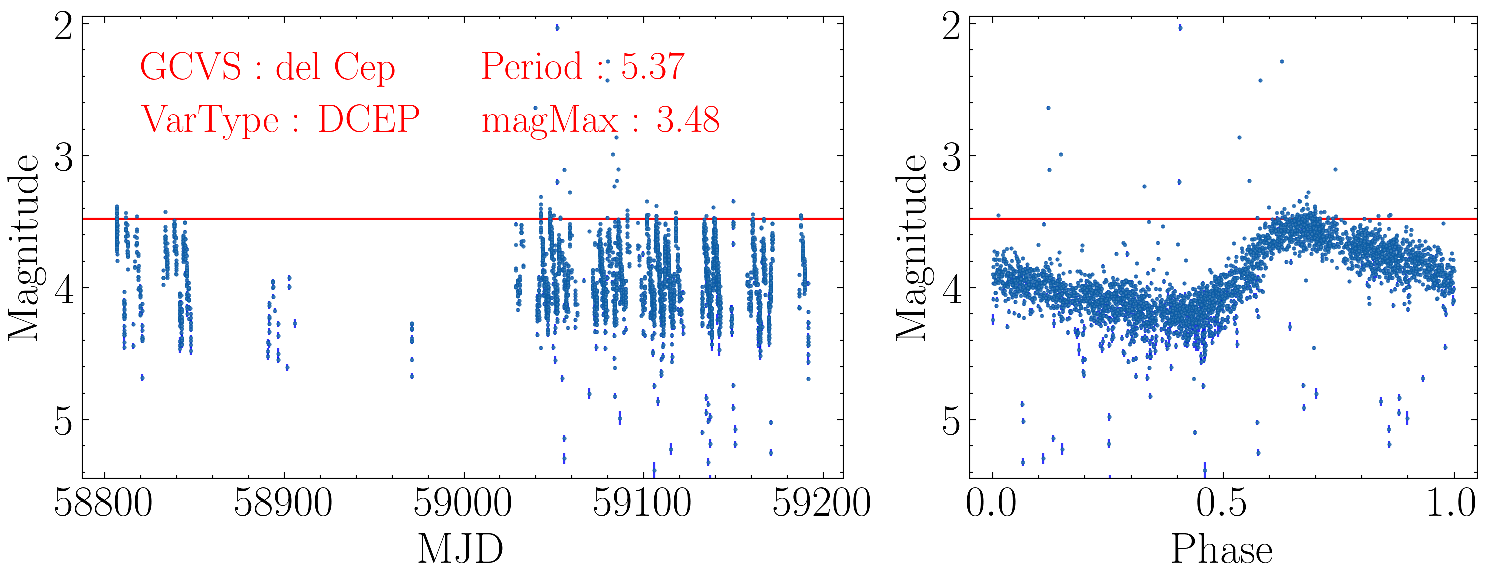}
\includegraphics[width=0.45\textwidth]{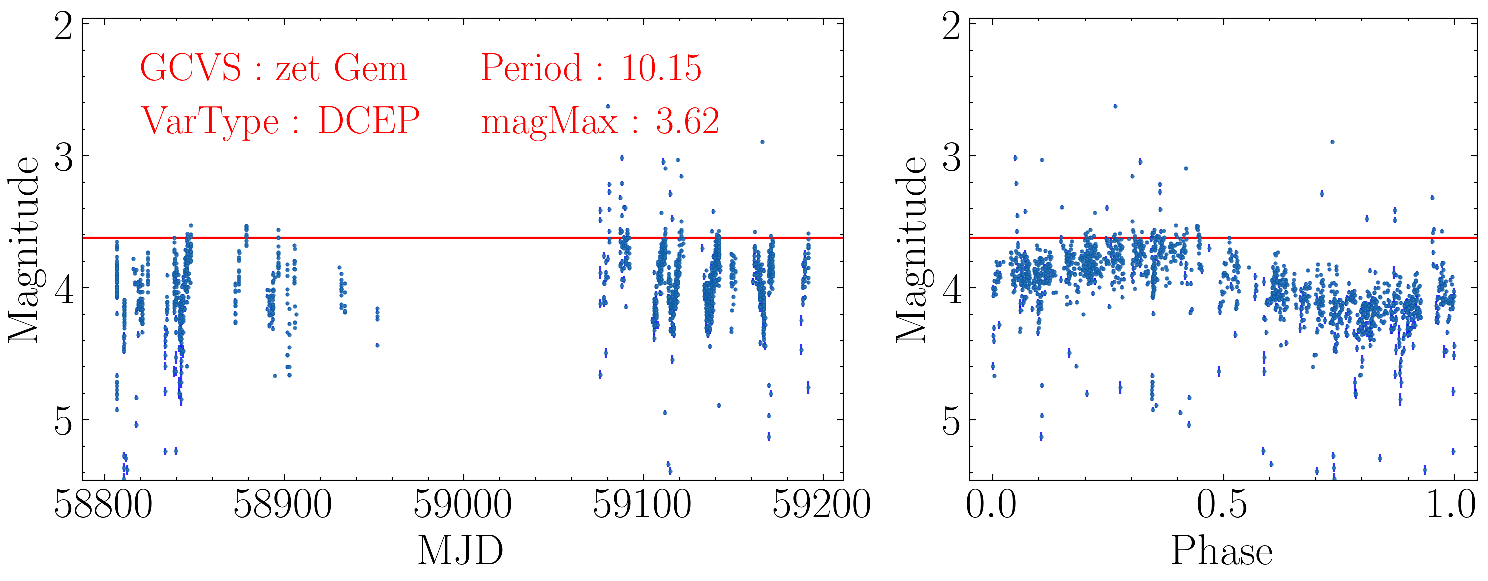}
$\quad$
\includegraphics[width=0.45\textwidth]{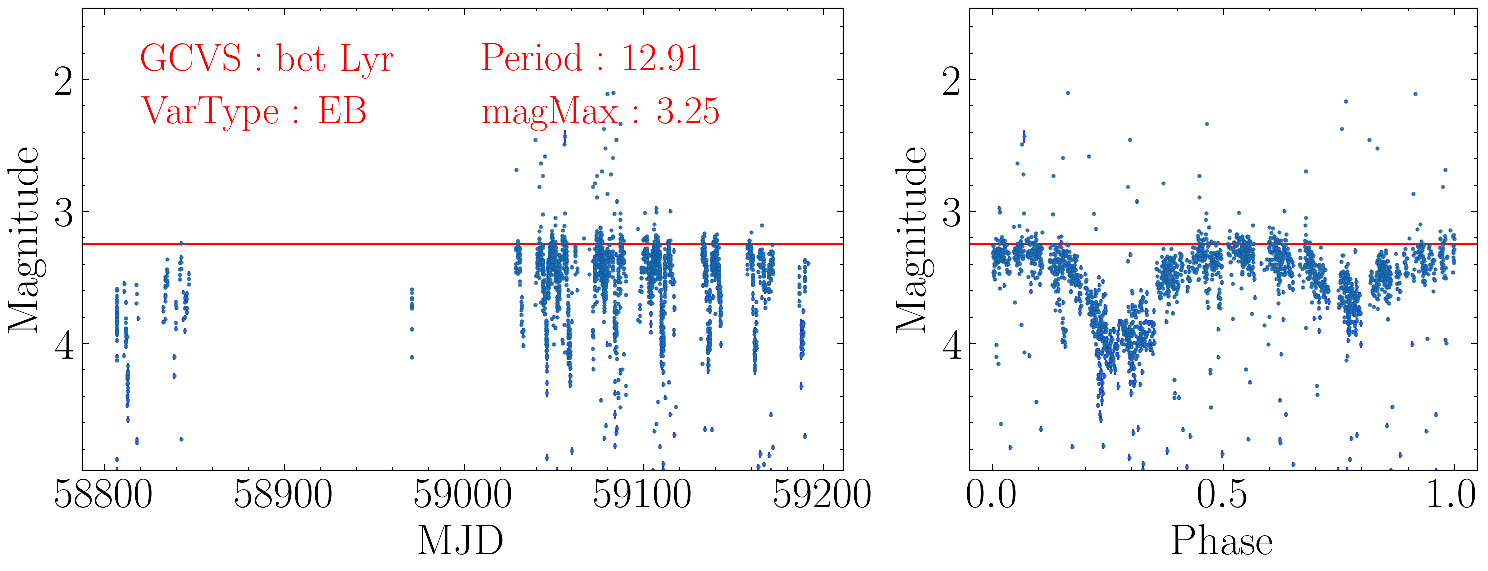}
\includegraphics[width=0.45\textwidth]{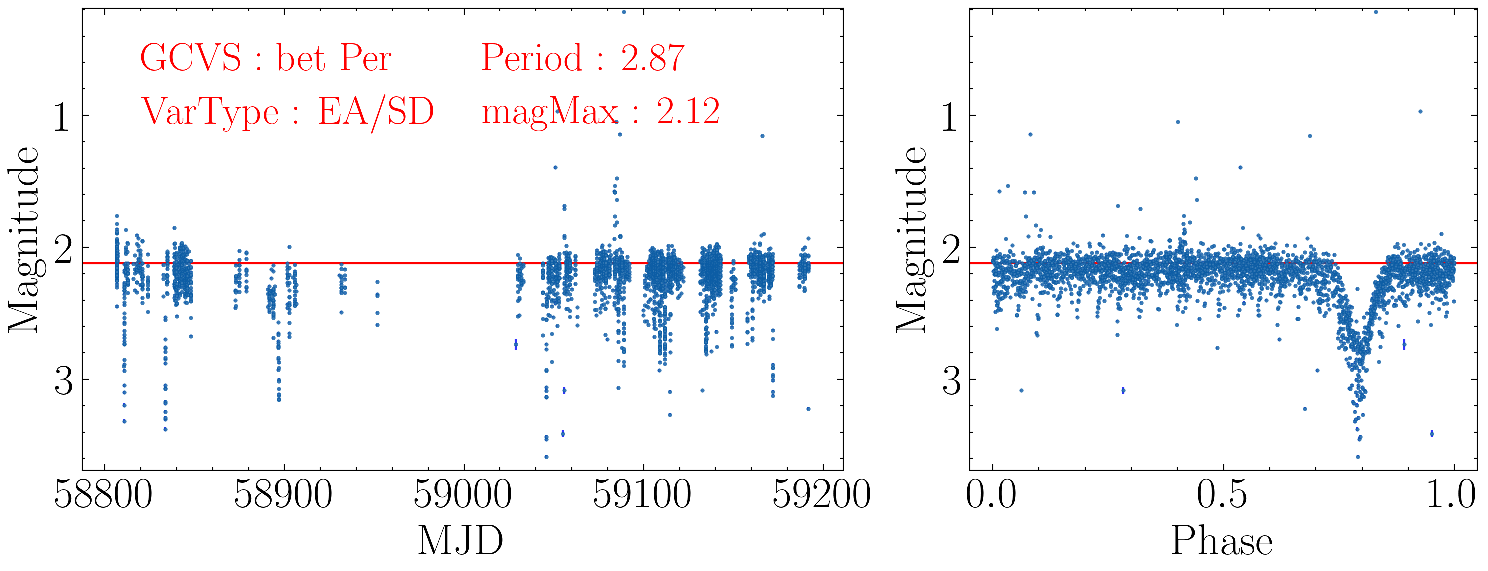}
$\quad$
\includegraphics[width=0.45\textwidth]{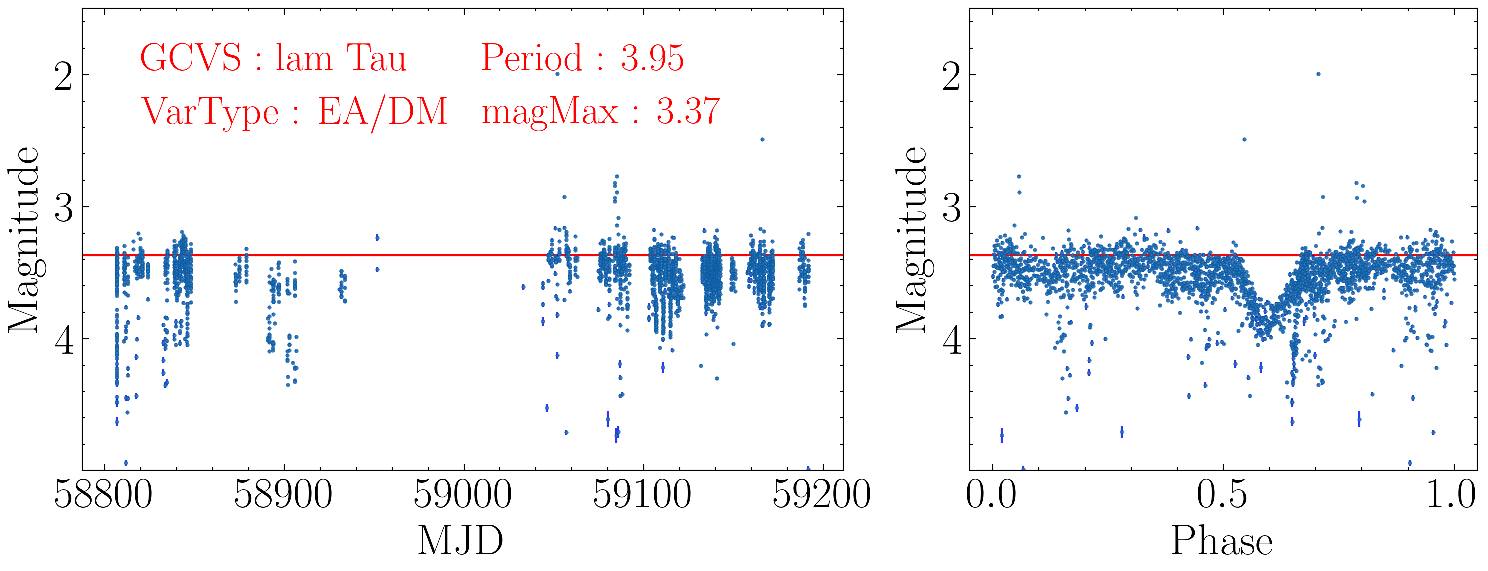}
\caption{The observed light curves of the known GCVS variable stars.}
\label{fig:gcvs_var}
\end{figure*}

To mitigate the impact of systematic errors and environmental variations during the observation period on the photometry results, calibration was performed using standard stars within five degrees of Polaris with magnitudes less than five, sourced from the GSC2 catalog \citep{2008AJ....136..735L}. The information of the four standard stars is presented in Table~\ref{tab:standard}, and their photometry variations are depicted in Fig.~\ref{fig:gsc2}. The graph illustrates that they exhibit significant systematic variations, and the patterns of change are consistent. After correcting for the systematic variations of the variable stars, we can derive their light curves. Fig.~\ref{fig:gcvs_var} presents the light curves of six known variable stars with significant amplitudes, whose variability characteristics are consistent with known information.

\begin{table}
\caption{The standard stars around Polaris.}
\label{tab:standard}
\begin{tabular}{lrrc}
\hline
GSC2 & RA & DEC & V\\
 & deg & deg & mag \\
\hline
NA9Z000575	&  17.19 & 86.26 & 4.24 \\
NA9S000157	&  37.95 & 89.26 & 2.04 \\
NA9S002719	&  37.97 & 89.26 & 2.00 \\
N3YH000343	& 263.05 & 86.59 & 4.35 \\
\hline
\end{tabular}
\end{table}

\section{Conclusions} \label{sec:conclusions}
The calibration model for an all-sky camera can be distilled into a transformation from the world coordinate system to the image coordinate system. The process is complicated by the projection and distortion characteristics inherent to fisheye lenses. Common calibration models address this complexity by employing various forms of projection models and polynomial distortion terms. The GWCS with AIPS convention can realize coordinate calibration of all-sky cameras.

Using HEALPix for the partitioning and transformation of all-sky images, we have successfully implemented star registration across the entire sky using Astrometry.net, which has allowed us to compile a star catalog for all-sky images. Subsequently, image coordinate calibration can be accomplished with GWCS, achieving a calibration accuracy of approximately 0.3\,pixels.

The number of detectable stars in an image is a crucial factor affecting the calibration accuracy of all-sky images. Stars with zenith angles less than 70$^{\circ}$ are abundant and can provide ample constraints for the calibration model. In contrast, stars with zenith angles greater than 70$^{\circ}$ are less numerous due to atmospheric extinction, failing to offer strong enough constraints for the calibration model. By using star registration results from multiple images, we can increase the number of stars with zenith angles greater than 70 degrees, which significantly enhances the calibration accuracy in those regions.

Based on the high-precision calibration method for the all-sky camera, we have accomplished the calibration of the cloud monitor at the Ali Observatory. We have measured the atmospheric extinction and night sky brightness at the observatory and obtained light curves for several known variable stars. The results indicate an average atmospheric extinction coefficient of 0.20 and a night sky background brightness of 21 mag arcsec$^{-2}$. There is evidence of light pollution in the direction of Shiquanhe City.

\section*{Acknowledgements}

This work is supported by National Key Research \& Development Program of China (No. 2021YFC2203202). We acknowledge support from the general grants 12273065, 11973004, 11873063 of the National Natural Science Foundation of China. 

\section*{Data Availability}

Due to specific national policy restrictions, the raw data of this study cannot be made publicly available. However, detailed data descriptions can be provided upon reasonable request. Please contact the corresponding author for more information. The author has provided a reference example for all-sky image calibration at https://github.com/yinjia86/Allskyimage.

\bibliographystyle{mnras}
\bibliography{cloudmethod}

\bsp	
\label{lastpage}
\end{document}